\documentclass[12pt,preprint]{aastex}

\usepackage{graphicx}
\usepackage{epstopdf}
\usepackage{url}
\usepackage{amsmath}
\usepackage{rotating}
\newcommand{\etaEarth}{\ensuremath{\eta_{\oplus}}}

\newcommand{\kepler}{\emph{Kepler}}

\newcommand{\rms}{\emph{rms}}

\newcommand{\svec}{\textit{\textbf{s}}}
\newcommand{\xvec}{\textit{\textbf{x}}}

\newcommand{\wvec}{\textit{\textbf{w}}}

\newcommand{\transpose}[1]{{#1}^{\rm{T}}}

\newcommand{\svecTilde}{\tilde{\svec}}
\newcommand{\wvecTilde}{\tilde{\wvec}}

\slugcomment{Submitted to the PASP.}

\shorttitle{Kepler Combined Differential Photometric Precision}
\shortauthors{Christiansen et al.}

\begin{document}


\title{The Derivation, Properties and Value of Kepler's Combined Differential Photometric Precision}

\author{Jessie L. Christiansen$^1$}
\author{Jon M. Jenkins$^1$}
\author{Thomas S. Barclay$^1$}
\author{Christopher J. Burke$^1$}
\author{Douglas A. Caldwell$^1$}
\author{Bruce D. Clarke$^1$}
\author{Jie Li$^1$}
\author{Shawn Seader$^1$}
\author{Jeffrey C. Smith$^1$}
\author{Martin C. Stumpe$^1$}
\author{Peter Tenenbaum$^1$}
\author{Susan E. Thompson$^1$}
\author{Joseph D. Twicken$^1$}
\author{Jeffrey Van Cleve$^1$}

\email{jessie.l.christiansen@nasa.gov}
\affil{$^1$SETI Institute/NASA Ames Research Center, M/S 244-30,  Moffett Field, CA 94035}




\begin{abstract}

The \kepler\ Mission is searching for Earth-size planets orbiting solar-like stars by simultaneously observing $>$160,000 stars to detect sequences of transit events in the photometric light curves. The Combined Differential Photometric Precision (CDPP) is the metric that defines the ease with which these weak terrestrial transit signatures can be detected. An understanding of CDPP is invaluable for evaluating the completeness of the \kepler\ survey and inferring the underlying planet population. This paper describes how the \kepler\ CDPP is calculated, and introduces tables of \rms\ CDPP on a per-target basis for 3-, 6-, and 12-hour transit durations, which are now available for all \kepler\ observations. Quarter~3 is the first typical set of observations at the nominal length and completeness for a quarter, from 2009 September 18 to 2009 December 16, and we examine the properties of the \rms\ CDPP distribution for this data set. Finally, we describe how to employ CDPP to calculate target completeness, an important use case.

\end{abstract}


\keywords{techniques: photometric --- methods: data analysis --- missions: Kepler}



\section{Introduction}
\label{s:intro}

The \kepler\ Mission is a NASA Discovery mission designed to detect transiting extrasolar planets, performing near-continuous photometric observations of $>$160,000 carefully selected target stars in \emph{Kepler's} 115 square degree field of view, as reviewed in \citet{Borucki2010} and \citet{Koch2010}. Scores of planets have been confirmed thus far{\footnote{See \url{http://kepler.nasa.gov/Mission/discoveries/}}, and three catalogues of planet candidates have been released: 705 candidates discovered in the first month of observations \citep{Borucki2011a}, 1235 candidates discovered in the first 15 months of observations \citep{Borucki2011b}, and 2321 candidates discovered in the first 18 months of observations \citep{Batalha2012}. Although individual planetary systems continue to surprise and intrigue---see for example the Kepler-36 system \citep{Carter2012}---we are now able to shift towards broader analysis of the underlying planetary populations \citep{Borucki2011a,Howard2011,Youdin2011,Catanzarite2011} and trends, such as comparing single planet candidate systems to those with multiple planet candidates \citep{Latham2011}, and the environments of small planet candidates compared to large planet candidates \citep{Buchhave2012}. 

The primary goal of the \kepler\ Mission is to ascertain the value of \etaEarth, the frequency of Earth-size planets orbiting in the habitable zones of solar-like stars. Inferring the value of \etaEarth\ from the planet sample discovered by \kepler\ requires careful quantification of the detectability of each planetary candidate across the entire set of target stars. An essential aspect of measuring \etaEarth\ is accounting for the observation noise specific to each target star, and the subsequent impact on the detectability of the transit signature of the candidate, which is the topic of this paper. 

\kepler's transiting planet search (TPS) pipeline module \citep{Jenkins2010,Tenenbaum2012}, which searches through the data for evidence of transit signatures, empirically determines the level of non-stationary noise for each light curve. This noise estimate is called the combined differential photometric precision (CDPP); it is a time series of the effective white noise as seen by a specific transit duration for each target star. To facilitate analysis and interpretation of the \kepler\ data, tables of the rms CDPP metrics on a per-target basis for transit durations of 3, 6, and 12 hours are provided online at the Mikulski Archive for Space Telescopes (MAST) website{\footnote {\tt http://http://archive.stsci.edu/kepler/}. Section~\ref{sec:cdppcalculation} of this paper describes how CDPP is calculated by the \kepler\ pipeline. Section~\ref{sec:tabledescription} provides a guide to the format and content of the tables. Section~\ref{sec:q3cdpp} discusses the general characteristics of the Q3 rms CDPP values, and examines the distribution as a function of stellar type and position in the \kepler\ field. Section \ref{sec:uses} describes some of the ways in which CDPP can be used for further analysis, and Section \ref{sec:conclusion} contains the conclusions of the paper.

\section{How CDPP is Calculated}
\label{sec:cdppcalculation}

The details of the pipeline processing prior to the CDPP calculation for a single quarter of data have been described in detail elsewhere; for an overview see \citet{Jenkins2010b}. Briefly, they involve pixel-level calibrations, including bias and dark current subtraction, flatfielding, and shutterless readout smear correction by the Calibration (CAL) module \citep{Quintana2010}; cosmic ray correction, background subtraction and simple aperture photometry by the Photometric Analysis (PA) module \citep{Twicken2010a}; systematic error removal and crowding correction of the flux time series by the Presearch Data Conditioning (PDC) module (Smith et al. 2012, Stumpe et al. 2012 for Release 12 onwards{\footnote{A data release is the delivery to MAST of a set of data products produced in a given processing run.}}); and finally, harmonic removal of strongly sinusoidal variations and flux time series extension (the latter for efficient fast Fourier transforms) by the Transiting Planet Search (TPS) module \citep{Jenkins2010}. When searching multiple quarters of data, TPS has several additional steps: median-normalize the flux level from quarter to quarter; fill the gaps between quarters (to enable the wavelet-based detection method described below); and detrend the discontinuities at the quarter boundaries \citep{Jenkins2010}. The signal detection is performed, generating threshold crossing events (TCEs) for further analysis; for example, \citet{Tenenbaum2012} presents the set of TCEs produced by TPS in the Q1--Q3 observations. We define a threshold crossing event as a signal which, when folded at a given period, gives rise to a signal of $\ge 7.1 \sigma$.

CDPP is calculated as a by-product of TPS, when determining the SNR of each transit pulse for which we search. Simply stated, the CDPP produced by TPS can be thought of as the effective white noise seen by a transit pulse of a given duration. A CDPP of 20~ppm for 3--hour transit duration indicates that a 3--hour transit of depth 20~ppm would be expected to have an SNR of 1, and hence, produce a signal of strength $1\sigma$ on average. Thus, CDPP is a characterization of the noise in the \kepler\ data.



Typically, the noise is non-white (that is, does not have a uniform power spectral density distribution) and non-stationary (that is, the power spectral density changes with time). The noise is typically dominated by $1/f$-type noise processes (where $f$ is frequency) due to stellar variability and instrumental effects; for a thorough breakdown and discussion of the noise sources contributing to CDPP see \citet{Gilliland2011} and \citet{VanCleve2009}. Therefore, we need a way of characterising the noise in the data in a moving fashion in order to preserve the time-variability; we achieve this by decomposing the data in the time-frequency domain using the wavelet approach. The theoretical basis of the approach is described in \citet{Jenkins02}. We measure the time-varying noise in a set of time-frequency bands, with equal spacing in log($f$), and use this estimate to adjust the noise level in each band to produce a `flat' power spectrum; that is, we filter the data to produce white noise. For consistency, the resulting whitening filter is also applied to the trial transit signal in order to reproduce and match any distortion created by the whitening. The detection of the whitened transit signal in the whitened data is therefore simplified to the well-understood problem of detecting a signal in the presence of white noise, and CDPP is a measure of the noise in the whitened data.

Full details of the calculation of CDPP in the SOC pipeline are given in \citet{Jenkins2010}; we briefly describe it here. For a given time series, $x(n)$, which is composed of a zero-mean, Gaussian noise process $w(n)$, with a power spectrum $P\left(\omega\right)$ and a corresponding autocorrelation matrix $R$, we want to detect a signal, $s(n)$ (a transit pulse in the context of \kepler). CDPP, the noise seen by the transit pulse, is quoted in ppm (parts per million), and is calculated from the detection statistic, $l$, by: 

\begin{equation}
\rm{CDPP} = 1 \times 10^6 / \left< \it{l} \right>
\label{eq:cdpp}
\end{equation}

The detection statistic is defined as:

\begin{equation}
	l = \frac{\xvec^{\rm{T}} ~R^{-1} \svec}{\sqrt{\svec^{\rm{T}}
	~R^{-1} \svec}} =
	\frac{\tilde{\xvec}^{\rm{T}}
	\tilde{ \svec}}
	{\sqrt{\tilde{\svec}^{\rm{T}} \tilde{\svec}}},
\label{eq:matchfilter}
\end{equation}

\noindent where $\tilde{\xvec} = R^{-\onehalf}\xvec$ and $\tilde{\svec} = R^{-\onehalf}\svec$ are data and signal vectors distorted or ``whitened" by the inverse square root of the autocorrelation matrix $R$. The expected value of the detection statistic, $\left<l\right>$, under the hypothesis (H1) that the signal $\svec$ is present, is given by:


\begin{align}
\label{eq:averageTestStatistic}
\notag \left<l\right>_{\rm{H1}} &= E\left\lbrace 
\frac{ \transpose{\left( \wvecTilde+\svecTilde \right) } \svecTilde} { \sqrt{\transpose{\svecTilde} \svecTilde}} 
\right\rbrace \\
&=
\frac{ E \left\lbrace \transpose{\wvecTilde} \svecTilde \right\rbrace +\transpose{\svecTilde} \svecTilde }  
{ \sqrt{\transpose{\svecTilde} \svecTilde}}= \frac{ 0 +\transpose{\svecTilde} \svecTilde }  
{ \sqrt{\transpose{\svecTilde} \svecTilde}}
\\
\notag &=\sqrt{\transpose{\svecTilde} \svecTilde} = \sqrt{\transpose{\svec}  ~R^{-1} \svec}.
\end{align}

For white noise, the auto-correlation matrix $R$ is diagonal, with $R = \sigma^2I$, where $\sigma^2$ is the variance of the white noise and $I$ is the identity matrix. Therefore, the mean value of the detection statistic in the presence of signal $s(n)$ is the signal-to-noise ratio (SNR) of the whitened signal to the whitened noise, as expected. Stated another way, the detection statistic is equivalent to measuring the significance of the signal from the chi-squared fit over the null hypothesis between the whitened light curve and the whitened transit signal. Under the null hypothesis (H0), $\left<l\right>_{\rm{H0}} = 0$. The two hypotheses have different mean values of $l$, but the same standard deviation of unity---this can be used to readily estimate the false alarm rate and detection rates, or their complements, using the error function. 

We calculate the time series of detection statistics $l$ for a given target over a set of trial durations; that is, we measure the detectability of given transit signals for each observation of each target. The CDPP time series is a natural by-product of this procedure, from Equation \ref{eq:cdpp}. We obtain CDPP on 14 time scales from 1.5--15 hours, which covers the transit durations of interest---for a central crossing event, an Earth-Sun analogue transit would take 13 hours, although for the average value of the impact parameter it would take 6.5 hours. The depth of an undiluted signal produced by such as system is 84~ppm: in a typical quarter of \kepler\ data, $12^{\rm{th}}$ magnitude dwarfs (the benchmark \kepler\ target) have a median rms CDPP value of $~$34~ppm for a transit duration of 6.5 hours. Although we produce a set of CDPP time series for each target, in practice we find that the amplitude of the variation in the CDPP over the time series for a given time scale is relatively small in a given quarter of data, and that to first-order, the detectability of a given transit signal is well described by the rms of the CDPP time series for that quarter. This motivated our production of the tables described in Section \ref{sec:tabledescription}.

\subsection{Some examples}
\label{sec:examples}

To illuminate the process described above, we show the light curves and resulting CDPP time series for several \kepler\ targets. The data presented are from the Q3 \kepler\ observations.

Figure \ref{fig:normalTarget} shows the detrended flux time series for target KIC~9392416, with magnitude Kp$=$11.7, and a 6-hour rms CDPP of 56~ppm. The power spectrum of this target is plotted as the red dashed line in Figure \ref{fig:powerSpectrum}, and we see that it is relatively uniform over all timescales, especially less than one day (i.e. the dominant noise source is white noise). An increase in noise in the flux time series can be seen at approximately days 113--114. The CDPP time series that are derived from this flux time series for transit pulse durations of 3, 6 and 12 hours are shown in Figure \ref{fig:normalTargetCDPP}. The aforementioned increase in noise at around days 113--114 manifests here as an increase in CDPP in the 3-hour time series at the same epoch. For the longer transit durations, the increased noise is averaged out and is not so apparent in the resulting 6-hour and 12-hour CDPP time series. This effect is also evident when considering the CDPP time series on the whole---note the decrease in the magnitude and flattening of the CDPP time series with increasing transit duration. The rms CDPP for this target decreases from 72~ppm for the 3-hour transit duration time series to 43~ppm for the 12-hour time series, as the signal is integrated over longer time spans.

\begin{figure}
\plotone{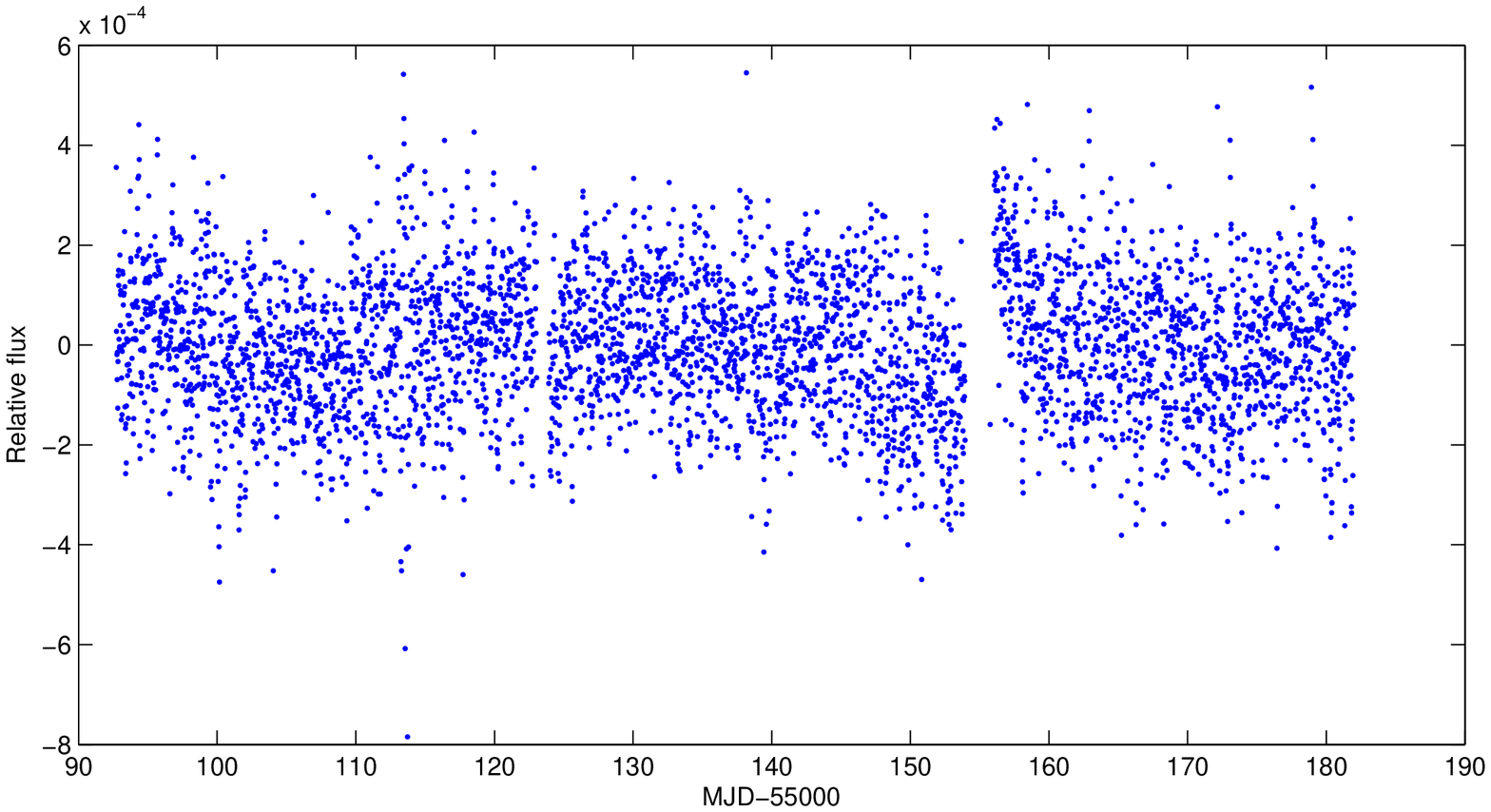}
\caption{The Q3 detrended flux time series for KIC~9392416, covering 89 days of observation. This target has a relatively uniform noise power spectrum.}
\label{fig:normalTarget}
\end{figure}

\begin{figure}
\plotone{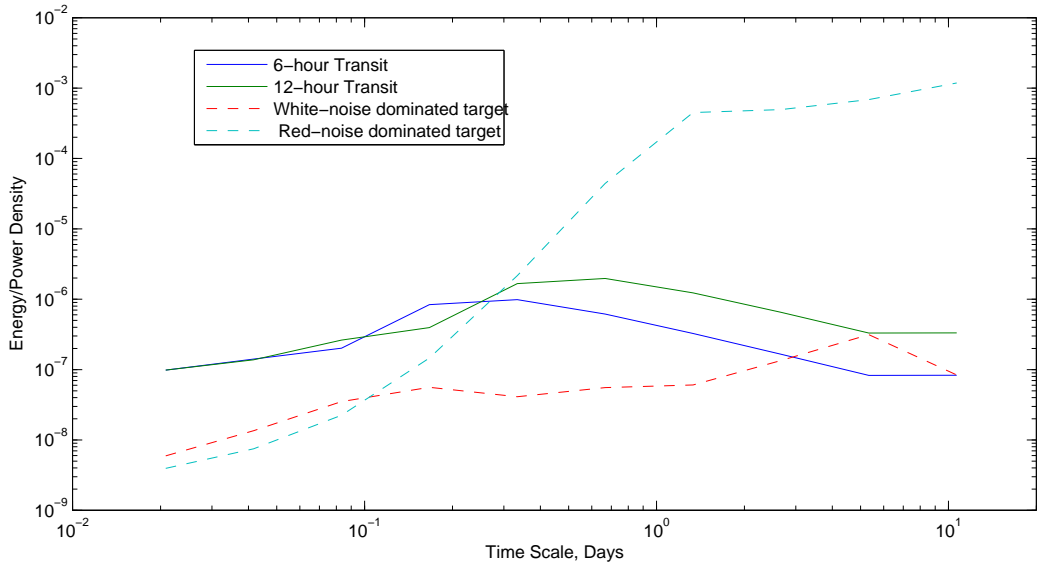}
\caption{A comparison of the noise power spectra of two \kepler\ targets with similar Kp magnitudes. The red dashed line is KIC~9392416 (see Figure \ref{fig:normalTarget}), which has a 6-hour rms CDPP of 56~ppm, Kp$=$11.6, and a roughly uniform power spectrum. The blue dashed line is the variable star KIC~9328434 (see Figure \ref{fig:variableTarget}), which has a 6-hour rms CDPP of 41~ppm, Kp$=$11.2, and a highly correlated power spectrum. Despite the significantly lower rms CDPP at approximately the same Kp magnitude, the variable target contains significantly more power at longer time scales. For comparison, we also show the energy over the same time scales for square pulse signals of 6- and 12-hours duration as blue and black solid lines respectively.}
\label{fig:powerSpectrum}
\end{figure}

\begin{figure}
\plotone{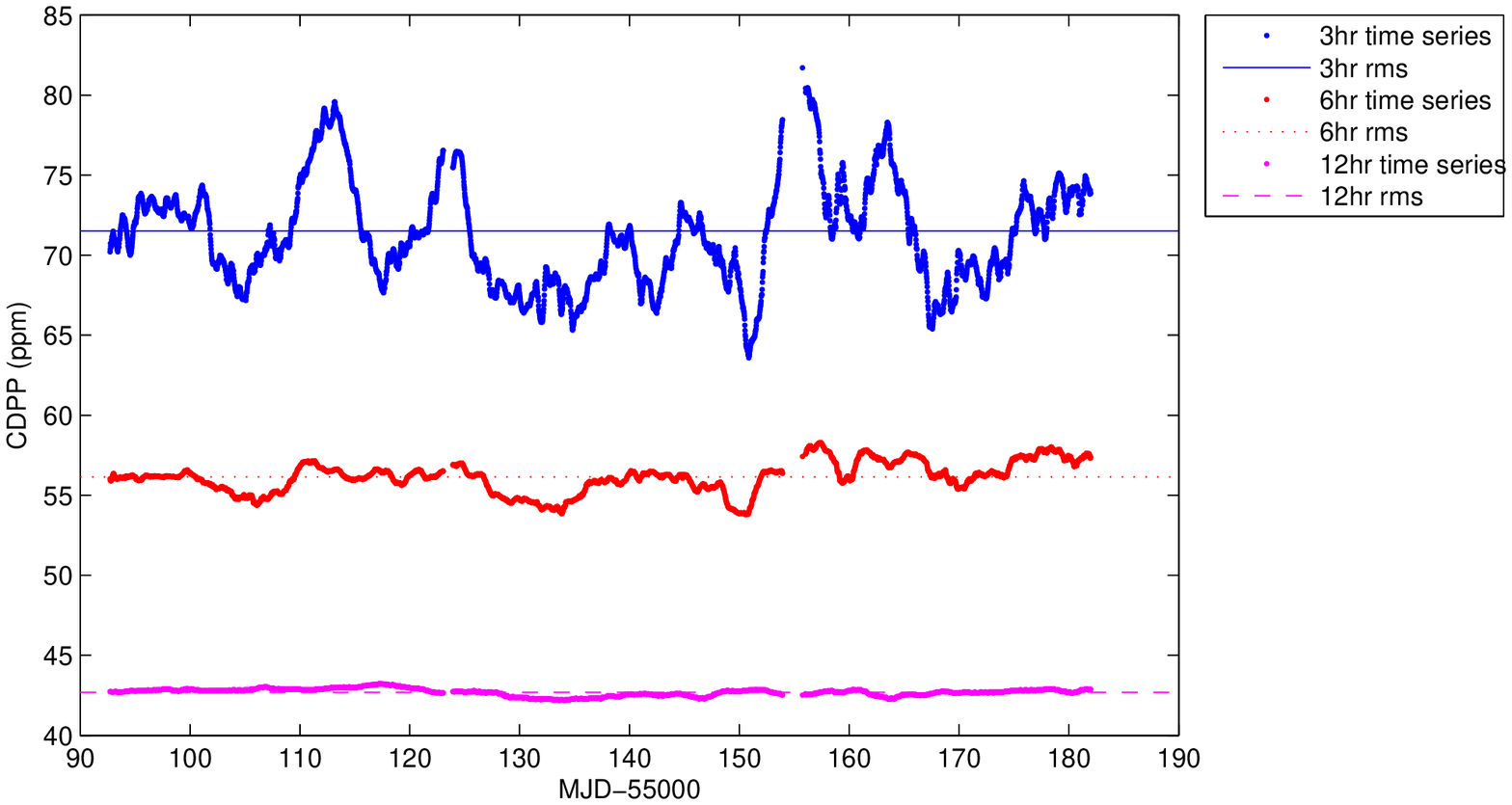}
\caption{The CDPP time series for KIC~9392416, shown for transit pulse durations of 3, 6, and 12 hours. The solid line shows the rms CDPP value for the 3-hour CDPP time series, the dotted line for the 6-hour CDPP time series, and the dashed line for the 12-hour CDPP time series.}
\label{fig:normalTargetCDPP}
\end{figure}

Figure \ref{fig:variableTarget} shows the detrended flux time series for KIC~9328434, which is a variable star with magnitude Kp$=$11.2. The blue dashed line in Figure \ref{fig:powerSpectrum} shows the power spectrum of this target, which is relatively quiet on short time scales (the 6-hour rms CDPP is 41~ppm), but contains significant correlated noise on longer time scales due to the stellar variability, which has a characteristic period of $\sim$2.5~days. Figure \ref{fig:variableTargetCDPP} shows the CDPP time series for this variable target. In this case, note that the 6-hour and 12-hour CDPP values are typically higher than the corresponding 3-hour CDPP. This arises, contrary to the previous, white-noise-dominated example, due to the fact that the multi-periodic flux time series has a characteristic period much longer than three hours. When integrating the flux time series on 6 and 12 hour time scales, we are integrating over larger changes in the stellar signal, which increases the scatter in the observations, and therefore the measured CDPP.

\begin{figure}
\plotone{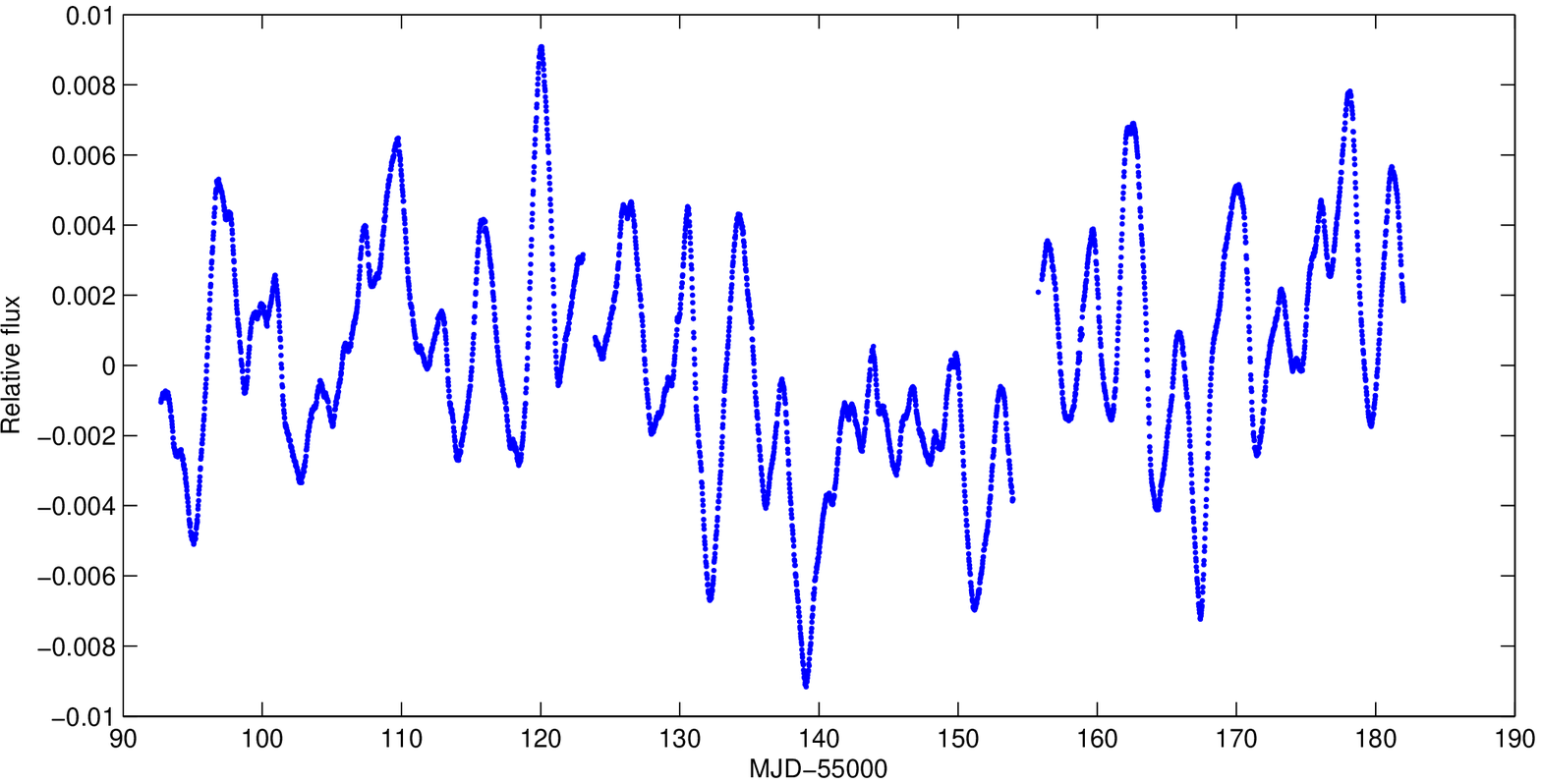}
\caption{The Q3 detrended flux time series for the variable star, KIC~9328434.}
\label{fig:variableTarget}
\end{figure}

\begin{figure}
\plotone{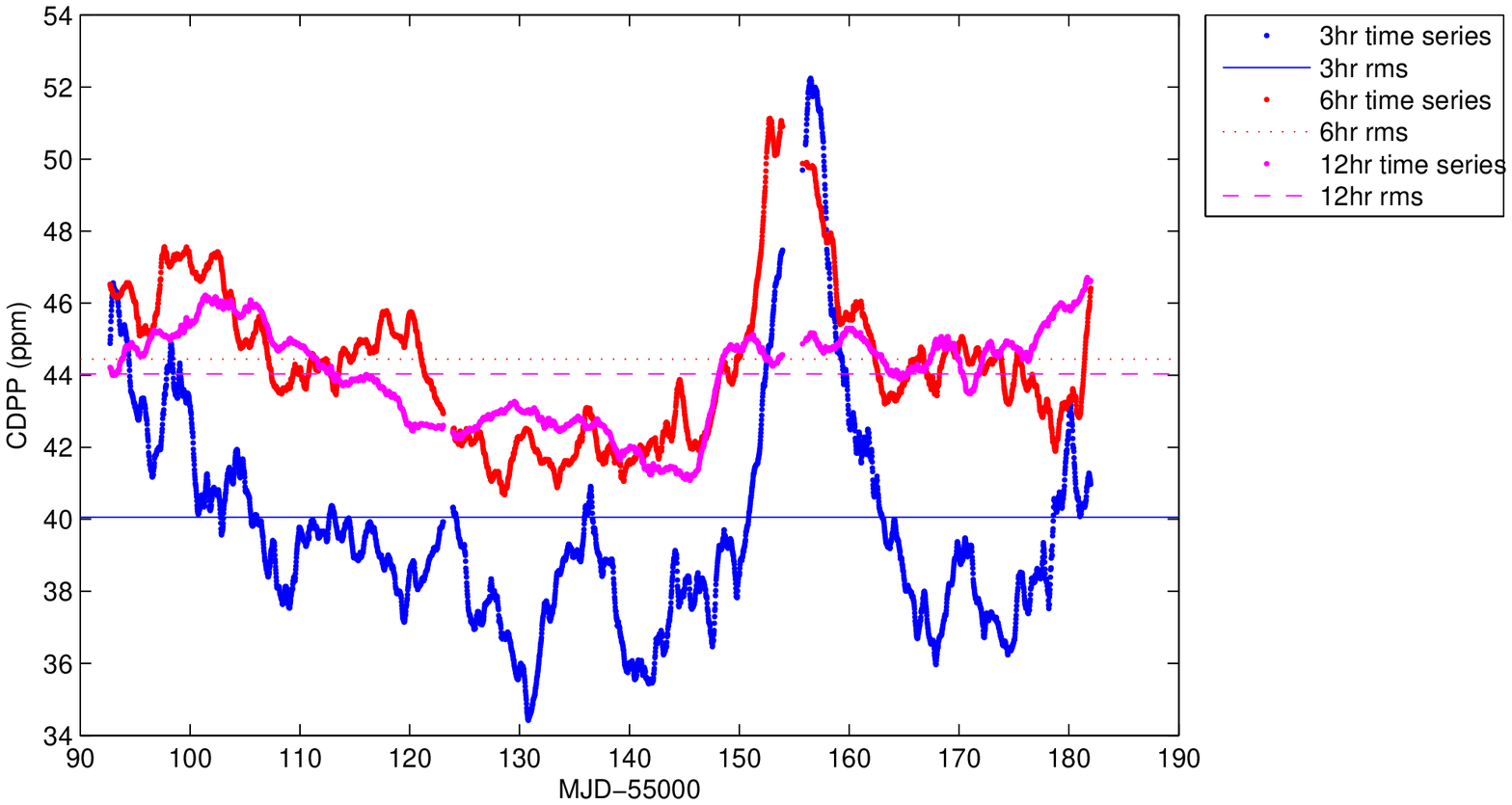}
\caption{The CDPP time series and rms CDPP for KIC~9328434, shown for transit pulse durations of 3, 6, and 12 hours.}
\label{fig:variableTargetCDPP}
\end{figure}

Finally, Figures \ref{fig:koi} and \ref{fig:koiCDPP} show the detrended flux time series and CDPP times series respectively for target KIC~9390653, which is also identified as KOI 249 in \citet{Borucki2011a}. This is a target with magnitude Kp$=$14.5, and a transiting planet candidate with an orbital period of 9.549 days. The detrended flux time series in Figure \ref{fig:koi} clearly shows the transits, which have an average depth of 1775~ppm. The transits have a duration of 1.84~hours, and although they have a high SNR, there is no evidence of their presence in the CDPP time series in Figure \ref{fig:koiCDPP}, that is, they do not perturb the locally measured noise; this is achieved by using a normalized median absolute deviation (MAD) rather than a true rms in the calculation of the in-band noise. Note that we also calculate CDPP time series for 1.5- and 2-hour transit pulses and the transits are not evident in these time series either; due to the nature of the whitening, transits should not generally increase the measured CDPP unless they are very closely spaced in time.

\begin{figure}
\plotone{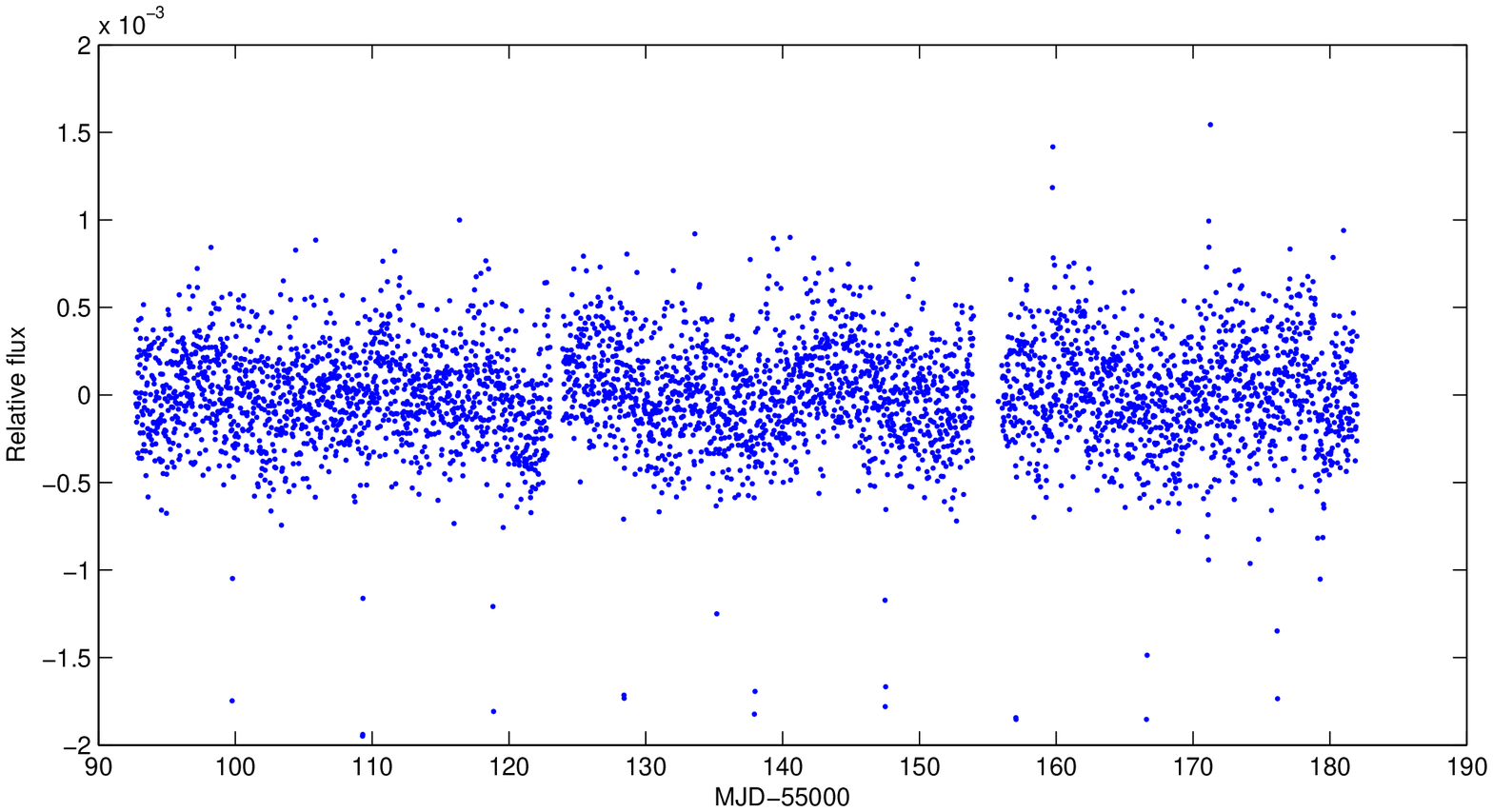}
\caption{The Q3 detrended flux time series for planet candidate KIC~9390653/KOI 249.}
\label{fig:koi}
\end{figure}

\begin{figure}
\plotone{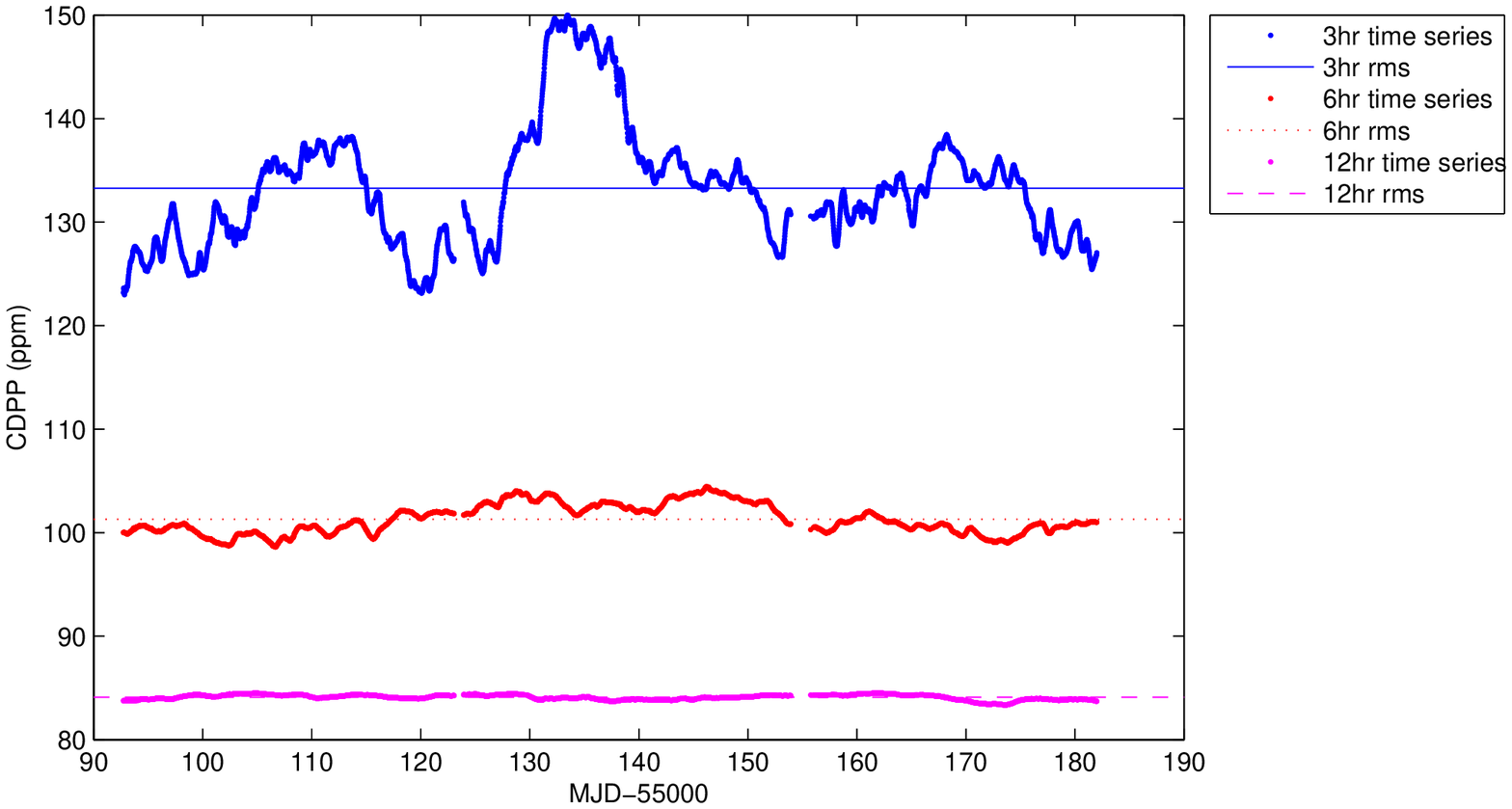}
\caption{The CDPP time series and rms CDPP for KIC~9328683, shown for transit pulse durations of 3, 6, and 12 hours.}
\label{fig:koiCDPP}
\end{figure}

\section{Description of online rms CDPP tables}
\label{sec:tabledescription}

For users interested in the ensemble noise statistics, for instance those performing completeness calculations, tables of the rms CDPP values for the planetary targets are being made available from the MAST \kepler\ website\footnote{For example \tt http://archive.stsci.edu/pub/kepler/catalogs/cdpp\_quarter3.txt.gz}, starting with all the publicly released data up to Quarter 10. In Table \ref{table:cdpp} we show an extract of the table for Quarter~3, described below, demonstrating the format and content. For users working with individual targets, the 3-, 6-, and 12-hour rms CDPP values are stored in the headers of the individual light curve FITS files, also available at the MAST website. As the data are reprocessed with improved versions of the SOC pipeline, the FITS files are replaced at MAST and will contain updated rms CDPP values. See the Kepler Archive Manual (KDMC-10008-004; Fraquelli \& Thompson 2012)\footnote{{\tt http://archive.stsci.edu/kepler/manuals/archive\_manual.pdf}} for details of the FITS headers.

\begin{sidewaystable}
\centering
\caption{An extract from the Q3 table of rms CDPP values available at the MAST website. Shown are the \kepler\ ID, the \kepler\ magnitude, the module and output on which the target is located in that quarter, the column and row location of the KIC position of the target, the fraction of the flux in the target aperture that is not from the target star, the stellar parameters from the KIC, including effective temperature ($T_{\rm eff}$), surface gravity (logg), stellar radius ($R_{\star}$), and colors (G-R and J-K); the data release number from the processing with which the CDPP values were calculated, the 3-hour, 6-hour and 12-hour rms CDPP values as described in the text, and the program ID of the target.}
\label{table:cdpp}
\footnotesize
\begin{tabular}{rrrrrrrrrrrrrrrrr}
\tableline\tableline
KIC & Kp & m & o & i & j & Contam & $T_{\rm eff}$ & logg & $R_{\star}$ & G-R & J-K & DRN & \multicolumn{3}{c}{rms CDPP} & PID \\
    &    &   &   &   &   &        &             &      &              &     &     &     & 3-hour    & 6-hour   & 12-hour &  \\
\tableline
   757076 & 11.678  & 16  & 4  &   27  &  106  & 0.020  &  5174 &  3.601 &   3.029 &  0.765 &  0.567 &    14 &    107.014 &   82.634 &   62.615 &  EX \\     
   757099 & 13.152  & 16  & 4  &   27  &  101  & 0.050  &  5589 &  3.817 &   2.288 &  0.648 &  0.563 &    14 &    620.885 &  515.472 &  424.305 &  EX \\      
   757137 &  9.196  & 16  & 4  &   19  &   71  & 0.002  &  4879 &  2.578 &  10.242 &  0.942 &  0.744 &    14 &     55.411 &   55.892 &   55.359 &  EX \\      
   757280 & 11.901  & 16  & 4  &   45  &   59  & 0.015  &  6648 &  4.082 &   1.683 &  0.341 &  0.225 &    14 &     53.880 &   45.545 &   40.292 &  EX \\      
   757450 & 15.264  & 16  & 4  &   77  &   52  & 0.138  &  5101 &   4.48 &   0.933 &  0.791 &  0.546 &    14 &    365.001 &  276.686 &  239.324 &  EX \\      
   891901 & 13.306  & 16  & 4  &   20  &  209  & 0.062  &  6051 &  4.411 &   1.085 &  0.419 &  0.286 &    14 &     83.043 &   62.044 &   50.276 &  EX \\      
   891916 & 14.799  & 16  & 4  &   21  &  206  & 0.184  &  5407 &  4.591 &   0.834 &  0.585 &  0.440 &    14 &    269.777 &  220.659 &  224.819 &  EX \\      
   892010 & 11.666  & 16  & 4  &   25  &  182  & 0.022  &  4665 &  2.413 &  12.342 &  1.024 &  0.743 &    14 &     90.958 &   91.752 &   92.047 &  EX \\      
   892107 &  12.38  & 16  & 4  &   26  &  155  & 0.029  &  5029 &  3.355 &   4.118 &  0.845 &  0.577 &    14 &    106.053 &   80.336 &   62.534 &  EX \\      
   892195 & 13.757  & 16  & 4  &   39  &  148  & 0.090  &  5553 &  3.972 &   1.865 &  0.704 &  0.502 &    14 &    114.754 &   84.893 &   64.444 &  EX \\    
\tableline
\end{tabular}
\end{sidewaystable}

\section{Quarter~3 rms CDPP values}
\label{sec:q3cdpp}

The summary statistics for the CDPP values are included in the Data Release Notes (DRN) for each quarter, available from the MAST website{\footnote {\tt http://archive.stsci.edu/kepler/datarelease.html}. Here we discuss in more detail the characteristics of the Q3 rms CDPP values, as a guide for analysis of subsequent quarters. Q3 was the first `typical' data set obtained by \kepler; the field was observed near-continuously from 2009 September 18 to 2009 December 16, for a total of 89 days. Thus it was the first opportunity to examine the distribution of rms CDPP values. Table \ref{tab:cdpp_by_mag} is an updated version of the Q3 values in Table 1 from the Kepler Data Release 14 Notes (KSCI-19054-001; Christiansen et al. 2012)\footnote{{\tt http://archive.stsci.edu/kepler/release\_notes/release\_notes14/DataRelease\_14\_2012032116.pdf}}, listing rms CDPP values for 6-hour transit durations instead of median CDPP values for 6.5-hour transit durations. The trends therein are discussed below.

\begin{table}[h!]
\begin{center}
\caption{Aggregate statistics for the Quarter~3 rms CDPP values for 6-hour transit durations. Column Definitions: (1) Kepler Magnitude at the center of the bin. Bins are ± 0.25 mag, for a bin of width 0.5 mag centered on this value. (2) Number of dwarfs (log $g > 4$) in the bin. (3) 10th percentile rms CDPP for dwarfs in the bin. (4) Median rms CDPP for dwarfs in the bin. (5) Number of all stars in the bin. (6) 10th percentile rms CDPP of all observed stars in the bin. (7) Median rms CDPP for all stars in the bin.}
\label{tab:cdpp_by_mag}
\begin{tabular}{|r|r|r|r|r|r|r|}
\tableline
\small{Kp mag} & \small{No. dwarfs} & \small{10th prctile} & \small{Median} & \small{No. stars} & \small{10th prctile} & \small{Median}\\
\tableline
   9.0 &     54   &   12.0 &  32.4 &     323 &  14.8 &  58.6 \\ 
  10.0 &    169   &   12.3 &  38.5 &     947 &  16.7 &  85.9 \\ 
  11.0 &    657   &   18.9 &  37.0 &    2573 &  22.7 &  78.2 \\ 
  12.0 &   2319   &   26.0 &  41.7 &    5810 &  28.9 &  70.7 \\ 
  13.0 &   7259   &   37.9 &  54.5 &   13003 &  39.9 &  67.6 \\ 
  14.0 &  14740   &   58.2 &  80.8 &   19420 &  59.2 &  86.2 \\ 
  15.0 &  27832   &  100.3 & 139.1 &   27904 & 100.2 & 139.1 \\ 
\tableline
\end{tabular}
\end{center}
\end{table}

\subsection{Distribution with stellar type}

Figure \ref{fig:q3cdppmag} shows the distribution of the 6-hour rms CDPP values with magnitude in the \kepler\ bandpass (Kp) for all the planetary targets in Q3, 165,441 targets in total. All stellar parameters discussed in this section are drawn from the \kepler\ Input Catalog (KIC; Brown et al. 2011). There are three distinct features visible in Figure \ref{fig:q3cdppmag}. The first is the discontinuity in the number of targets at Kp$=14$---this is an artifact of the \kepler\ target selection, whereby targets with Kp $>14$ and log $g \le 4$ are excluded to reduce the number of populous faint giants in the target list. The second is the increase in rms CDPP with increasing magnitude. The lower bound on this distribution is the minimum noise floor, with contributions from shot noise, which increases with increasing magnitude, and the typical read noise. The third is that there is a faint population separated vertically from the main body of targets. The difference in the populations can be attributed directly to the stars (i.e. it is not an instrumental effect). The upper panel of Figure \ref{fig:q3cdpp_dwarfs_giants} shows the distribution of rms CDPP for dwarf stars with log $g > 4$, and the lower panel shows the distribution for giant stars with log $g \le 4$ (note the cut-off at Kp$=14$, as stated previously). The bin sizes are 2~ppm for the rms CDPP and 0.05 magnitudes for the \kepler\ magnitude. The observed noise levels for the giant stars are significantly higher on average; in particular the population between 160--240~ppm which is virtually absent in the dwarf stars{\footnote The lower population will include sub-giants that are not fully evolved off the main sequence and possibly some number of stars with incorrect log $g$ values in the KIC.}. This was first noted for the \kepler\ targets by \citet{Koch2010}. This is also observed when plotting the rms CDPP values as shown in Figure \ref{fig:q3cdppHR}, with the surface gravity plotted against the effective temperature of the target. The giant stars, with low surface gravities, are distinct as the relatively high rms CDPP population in the top right of the figure. The coolest dwarf stars are also highly active, which increases the observed noise. These stars comprise another relatively high rms CDPP population in the bottom right. Solar-type stars, with $T_{\rm eff}\sim$ 5500K and log $g\sim$ 4, are typically well-behaved on the timescales over which we are measuring the noise, and comprise the relatively low rms CDPP population in the center. \citet{Batalha2012} noted that there are systematic problems with the stellar parameters in the KIC, with combinations of surface gravities and effective temperatures that do not lie on any modelled isochrone. As a result, there are likely to be significant numbers of misclassified stars in this figure (see \citet{Mann2012} and \citet{Brown2011} for more discussion on these biases), however the intent here is to demonstrate the behavior of CDPP as a broad function of stellar parameters, to allow users to investigate the parameter space for selecting targets for further analysis. 

Figure \ref{fig:cdpp_hist} shows the distribution of the rms CDPP values for the targets from the upper panel of Figure \ref{fig:q3cdpp_dwarfs_giants}, i.e. dwarf stars with log$g >$ 4, around which \kepler\ can expect to find transits of Earth-size planets; this is also the subset of targets considered in the `dwarfs' columns of Table \ref{tab:cdpp_by_mag}. Although bright targets are preferable for follow-up observations, the large increase in the number of \kepler\ targets with increasing magnitude is clear; in fact most \kepler\ planets will be found around stars fainter than Kp$=14$. The mode of the rms CDPP values increases with increasing magnitude largely due to increased contributions from shot noise.

\begin{figure}
\plotone{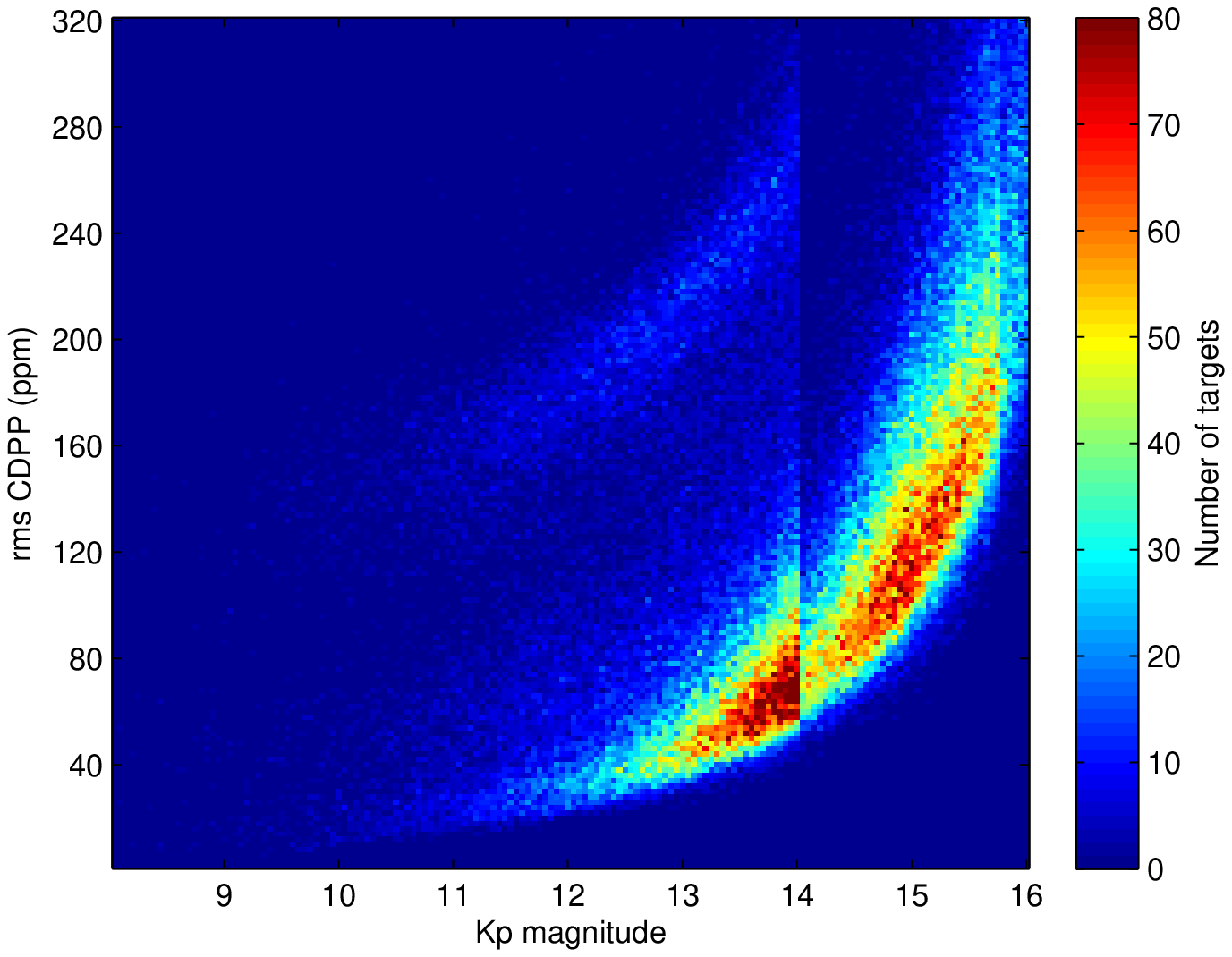}
\caption{The distribution of the 6-hour rms CDPP values with Kp magnitude for all Quarter~3 planetary targets.}
\label{fig:q3cdppmag}
\end{figure}

\begin{figure}
\plotone{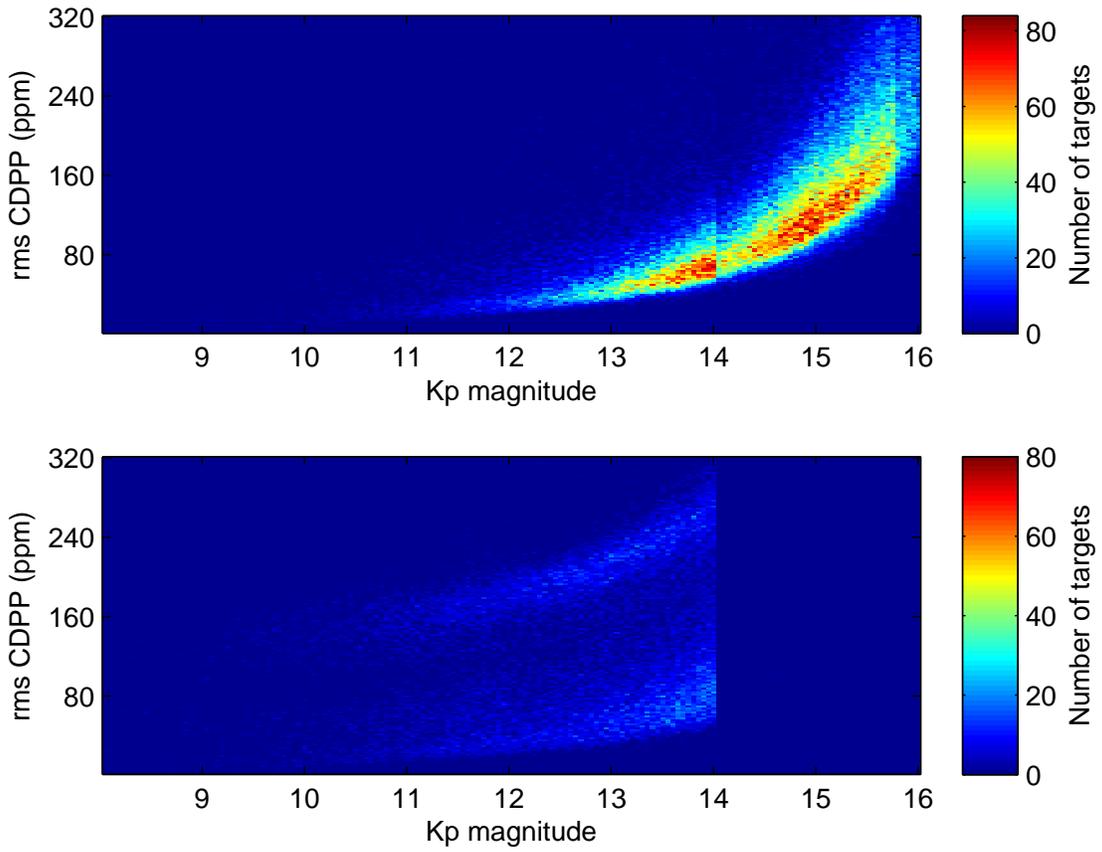}
\caption{The distribution of the 6-hour rms CDPP values with Kp magnitude for Quarter~3 dwarf targets (upper panel) and giant targets (lower panel).}
\label{fig:q3cdpp_dwarfs_giants}
\end{figure}

\begin{figure}
\plotone{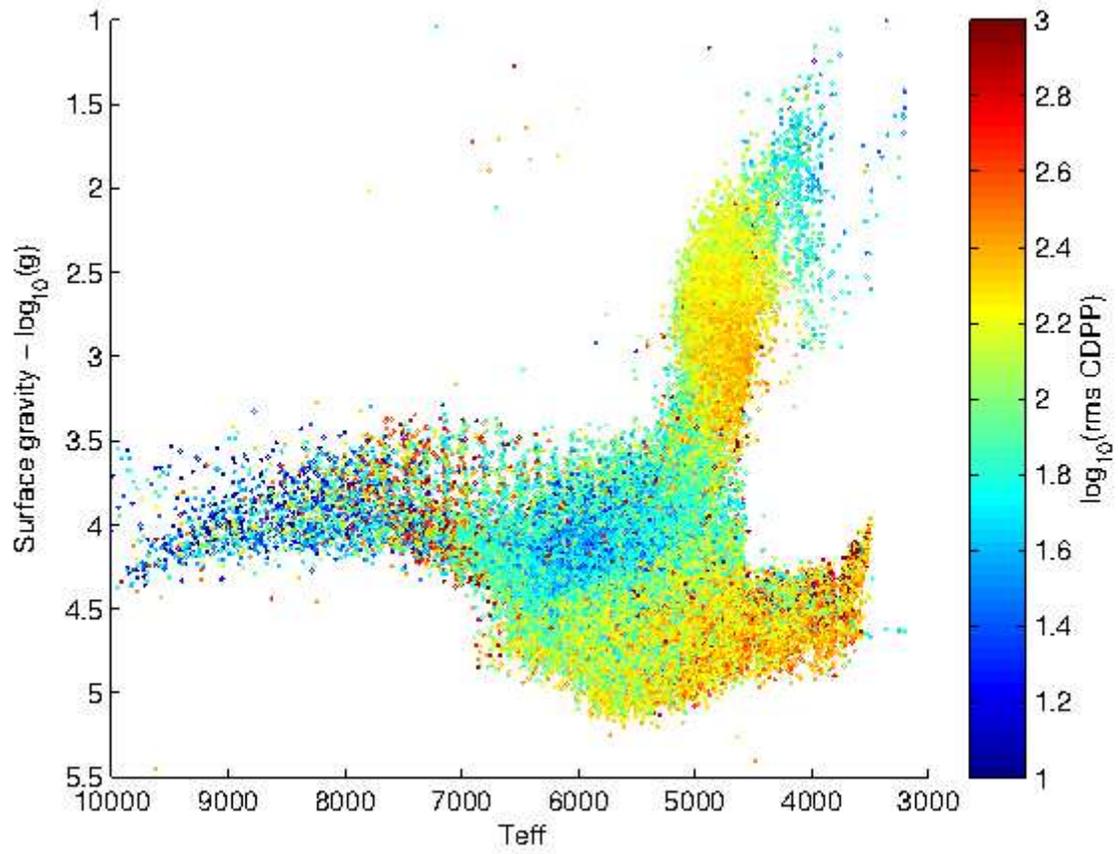}
\caption{The distribution of 6-hour rms CDPP values with KIC stellar parameters.}
\label{fig:q3cdppHR}
\end{figure}

\begin{figure}
\plotone{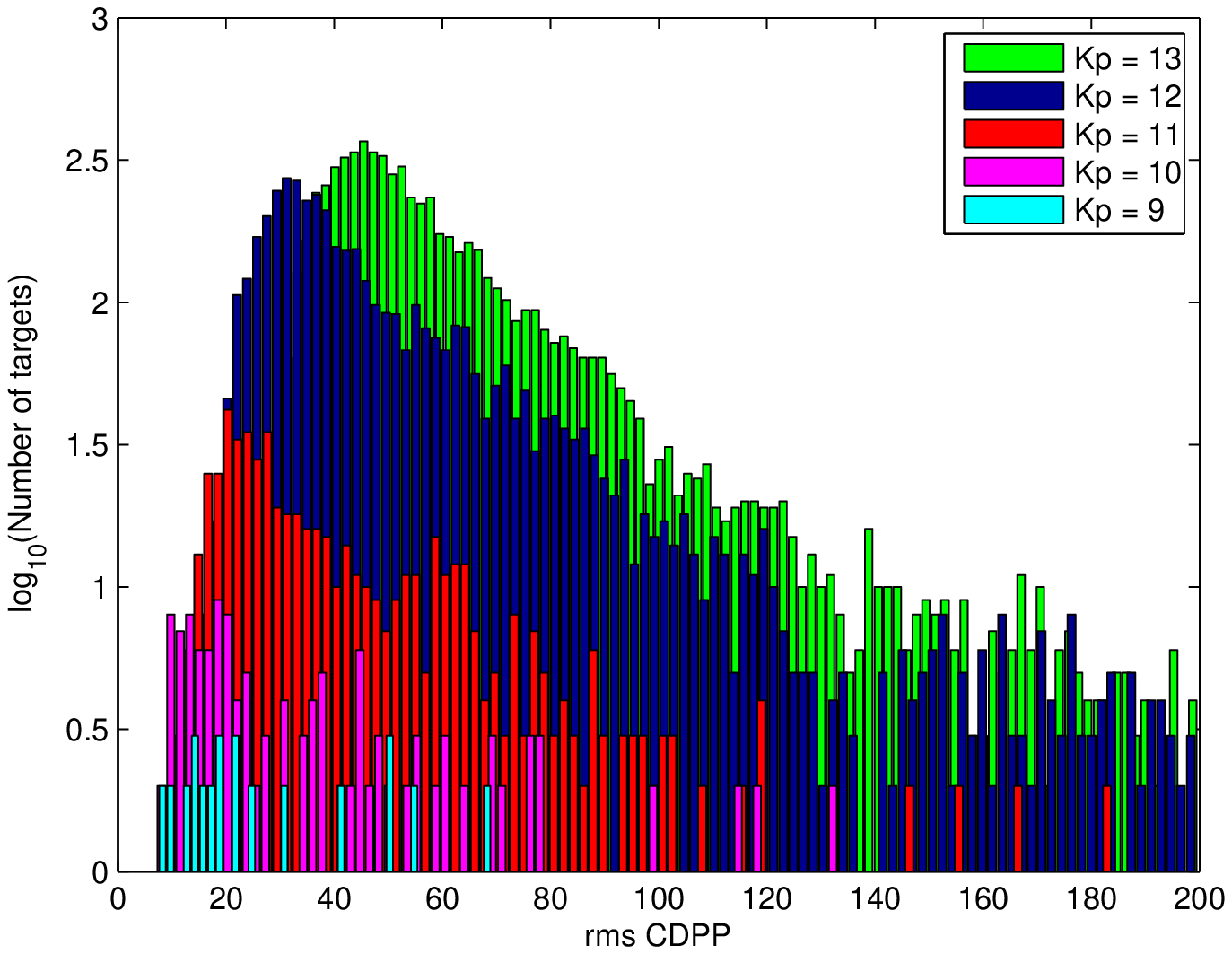}
\caption{The distribution of rms CDPP values for 6-hour transit durations in magnitude bins listed in Table \ref{tab:cdpp_by_mag}. These distributions are used for the rms CDPP statistics in that table.}
\label{fig:cdpp_hist}
\end{figure}

\subsection{Distribution with position in the field of view}

Figure \ref{fig:q3cdppradec} shows the distribution of the Quarter~3 rms CDPP values across the \kepler\ field of view, for the same targets as Figure \ref{fig:q3cdppHR}. The 21 modules can be seen projected onto the sky coordinates, each with two CCDs. The black arrow points in the direction of the Galactic plane. Although it is a fairly uniform distribution overall, two slight trends are evident. The first is a correlation between rms CDPP and the quality of the focus. In order to maximize the number of targets with good focus across the field, \kepler's best focus is found in the modules surrounding the central module. The central module itself and the outer modules are slightly out of focus compared to these modules. The modules with the best focus have lower rms CDPP values on average due to the resultant lessening of aperture effects introduced by pointing jitter and differential velocity aberration. The second trend is a function of Galactic latitude---modules closest to the Galactic plane have a slightly higher rms CDPP on average. This is most notable in the southern-most module in Figure \ref{fig:q3cdppradec}. This is a result of the increased stellar density toward the Galactic plane, which increases the amount of noise in a given pixel contributed by background stars relative to modules at higher Galactic latitudes.

\begin{figure}
\plotone{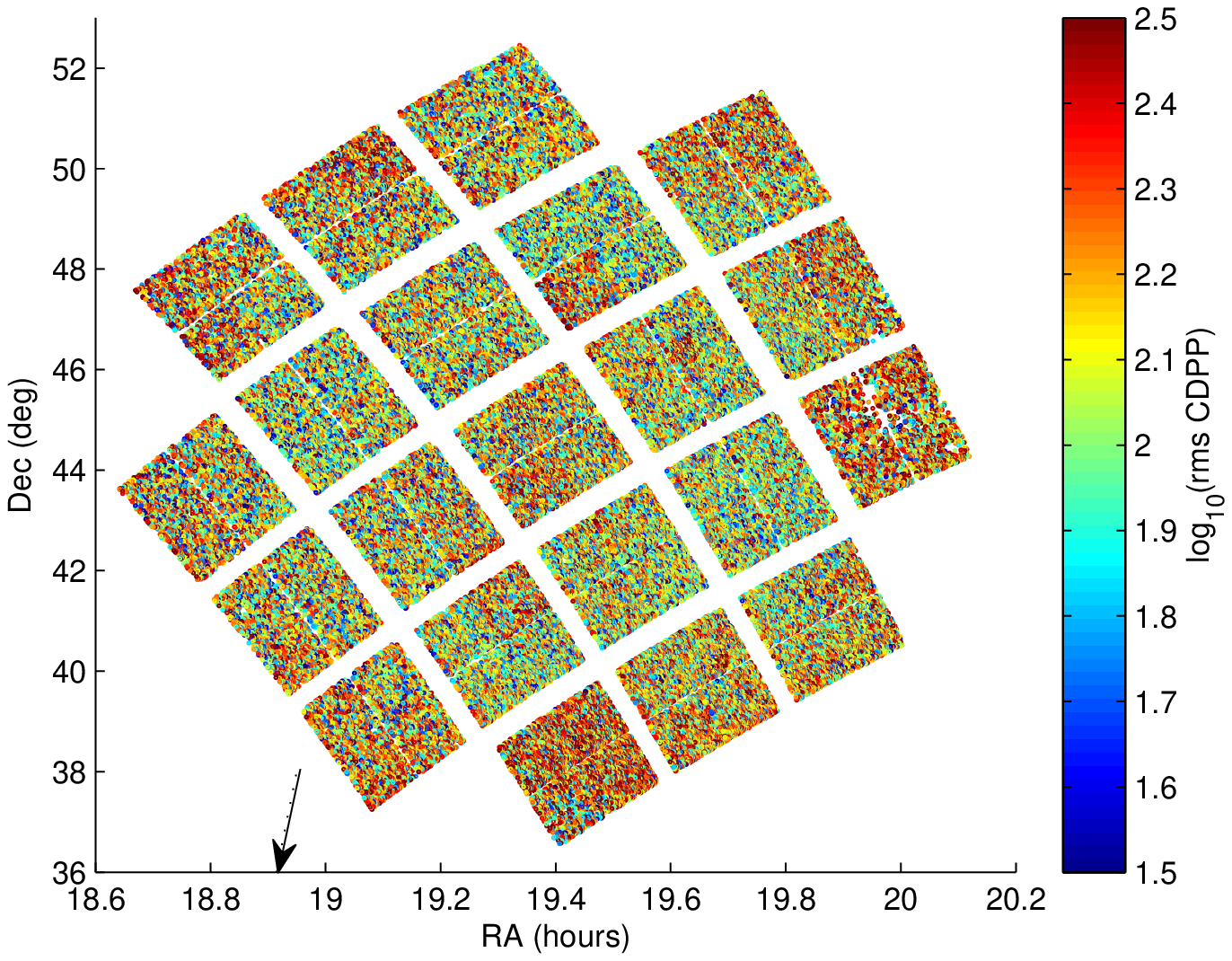}
\caption{The distribution of rms CDPP values on the \kepler\ field of view.}
\label{fig:q3cdppradec}
\end{figure}

\section{Using rms CDPP to estimate completeness}\label{sec:uses}

As described in Section \ref{sec:cdppcalculation}, CDPP is a direct, empirical measurement of the detectability of a given transit signature in the \kepler\ data. Using the CDPP time series for a target, it is possible to calculate the probability that a sample transit signal of a given depth and duration, occurring at a given time, could be detected by the pipeline. The rms of that CDPP time series represents, to first-order, the average detectability of that sample transit signal for that target.

For the simplest derivation, we set $t_{\rm obs}$ to the total span of time encompassed by observations of the target, and $f_{o}$ to the fraction of the total time the target was observed\footnote{The typical observation duty cycle for targets observed continuously is $\sim$92\% when accounting for gaps due to data downlink and spacecraft operations.}. For a given period, this gives us the average number of transits, $N_{tr}=(t_{\rm obs}*f_{o})/P$, observed for a signal with period $P$. The total SNR that would be measured for a planet at period $P$ over the whole time series is then $\sqrt{N_{tr}} \times \sigma_{1}$, where $\sigma_{1}$ is the SNR of a single transit event. In an assumed white noise regime, the effective rms CDPP, $CDPP_{\rm eff}$, for a given transit duration, $t_{\rm dur}$, can be estimated by finding the closest provided duration, $t_{\rm CDPP}$ (out of 3, 6 and 12 hours), and scaling the rms CDPP of that duration, $CDPP_{\rm N}$, such that $CDPP_{\rm eff}=\sqrt{t_{\rm CDPP}/t_{\rm dur}}\times CDPP_{\rm N}$. See the discussion in Section \ref{sec:examples} for a caution regarding applying this estimation to variable targets. For a planet of radius $R_p$, the transit depth is $\delta=(R_p/R_S)^2$ for a target star of radius $R_S$. For a single transit, $\sigma_{1}$ is then simply $\delta/CDPP_{\rm eff}$, and the SNR of a putative planet with period $P$ and radius $R_P$ over the set of observations is:

\begin{equation}
SNR = \sqrt{\frac{t_{\rm obs}*f_{o}}{P}}\frac{\delta}{CDPP_{\rm eff}}
\end{equation}

similarly to Equation 1 of \citet{Howard2011}.

In the \kepler\ pipeline, the SNR threshold for detection is 7.1$\sigma$ \citep{Jenkins2002b}. Using the above calculation, it is then possible to estimate whether the putative planet signal would have been detected by the pipeline. This estimation can be performed over a grid of planet parameters for a target star, for which the detection completeness can then be calculated. See Section 7.1 of \citet{Batalha2011} for a worked example of a typical $12^{\rm th}$ magnitude target star. For informed analyses of the planet population produced by \kepler\ thus far, the tables of rms CDPP are a necessary resource.

\section{Conclusion}\label{sec:conclusion}

We have presented here an introduction to the rms CDPP values being made available on a per-target, per-quarter basis at the MAST website for the \kepler\ planetary targets. For each set of data, these values provide a measure of the observed noise, which for each individual target translates to a limit on the detectability of transiting planets. These values are extremely important for the characterisation of the underlying planet population, since to first order, they provide on a star-by-star basis the observational noise level that sets the limiting planet signal detectable by Kepler.

\acknowledgments

Funding for the \kepler\ Discovery Mission is provided by NASA's Science Mission Directorate. We thank the thousands of people whose efforts made \emph{Kepler's} grand voyage of discovery possible. Some/all of the data presented in this paper were obtained from the Mikulski Archive for Space Telescopes (MAST). STScI is operated by the Association of Universities for Research in Astronomy, Inc., under NASA contract NAS5-26555. Support for MAST for non-HST data is provided by the NASA Office of Space Science via grant NNX09AF08G and by other grants and contracts.






\clearpage
\end{document}